\documentclass[sn-mathphys]{sn-jnl}
\usepackage{subfigure}
\usepackage{amsmath}
\usepackage{color}
\usepackage{soul}

\theoremstyle{thmstyleone}%
\theoremstyle{thmstyletwo}%

\theoremstyle{thmstylethree}%

\begin{document}

\title[Benchmarking the performance of portfolio optimization with QAOA]{Benchmarking the performance of portfolio optimization with QAOA}


\author[1]{\fnm{Sebastian} \sur{Brandhofer}}
\author[2]{\fnm{Daniel} \sur{Braun}}
\author[4]{\fnm{Vanessa} \sur{Dehn}}
\author[3]{\fnm{Gerhard} \sur{Hellstern}}
\author[2]{\fnm{Matthias} \sur{H\"uls}}
\author[1]{\fnm{Yanjun} \sur{Ji}}
\author[1]{\fnm{Ilia} \sur{Polian}}
\author[2]{\fnm{Amandeep Singh} \sur{Bhatia}}
\author*[4]{\fnm{Thomas} \sur{Wellens}}
\email{thomas.wellens@iaf.fraunhofer.de}

\affil[1]{\orgdiv{Institut für Technische Informatik}, \orgname{Universität Stuttgart}, \orgaddress{\street{Pfaffenwaldring 47}, \city{D-70569 Stuttgart}, \country{Germany}}}

\affil[2]{\orgdiv{Institut für Theoretische Physik}, \orgname{Eberhard Karls Universität Tübingen}, \orgaddress{\street{Auf der Morgenstelle 14}, \city{D-72076 T\"ubingen}, \country{Germany}}}

\affil[3]{\orgdiv{Zentrum für Digitale Innovationen (ZDI)}, \orgname{DHBW Ravensburg}, \orgaddress{\street{Marktstr. 28}, \city{D-88214 Ravensburg}, \country{Germany}}}

\affil[4]{\orgname{Fraunhofer-Institut f\"ur Angewandte Festk\"orperphysik IAF}, \orgaddress{\street{Tullastr. 72}, \city{D-79108 Freiburg}, \country{Germany}}}


\abstract{
We present a detailed study of portfolio optimization using different versions of the quantum approximate optimization algorithm (QAOA). For a given list of assets, the portfolio optimization problem is formulated as quadratic binary optimization constrained on the number of assets contained in the portfolio. QAOA has been suggested as a possible candidate for solving this problem (and similar combinatorial optimization problems) more efficiently than classical computers in the case of a sufficiently large number of assets. However, the practical implementation of this algorithm requires a careful consideration of several technical issues, not all of which are discussed in the present literature. The present article intends to fill this gap and thereby provide the reader with a useful guide for applying QAOA to the portfolio optimization problem (and similar problems). In particular, we will discuss several possible choices of the variational form 
and of different classical algorithms for finding the corresponding optimized  parameters. Viewing at the application of QAOA on error-prone NISQ hardware, we also analyze the influence of statistical sampling errors (due to a finite number of shots) and gate and readout errors (due to imperfect quantum hardware). Finally, we define a criterion for distinguishing between \lq easy\rq\ and \lq hard\rq\ instances of the portfolio optimization problem.}




\maketitle

\section{Introduction}
\label{sec1}

The solution of combinatorial optimization problems with QAOA, i.e., the quantum approximate optimization algorithm \cite{farhi14} 
or its generalized version known as quantum alternating operator ansatz \cite{hadfield19}, is often mentioned as one of the most promising candidates for realizing an advantage of quantum computers with respect to classical computers in a relatively near future. This expectation is based upon the fact the quantum circuits required for QAOA are shorter than those of other algorithms, e.g the Shor \cite{shor} or HHL \cite{hhl} algorithm, for which an exponential advantage compared to classical algorithms can be proven (under the assumption that quantum computers with sufficiently many error-corrected qubits are available).
In contrast to those cases, a quantum advantage of QAOA has not yet been rigorously proven. Since QAOA involves a  numerical classical optimization of variational circuit parameters (which, in itself, is an NP-hard problem \cite{bittel21}), analytical results concerning the performance of QAOA for a large number of qubits are not available. Nevertheless, extrapolations of results obtained for small numbers of qubits towards larger numbers  suggest that a quantum speed-up (for the Max-Cut problem) could be expected with several hundred qubits \cite{guerreschi19}. As pointed out there, the main limitation of QAOA is imposed by the exponential cost required for optimizing the variational parameters. 

This conclusion, however, might change if the exponential cost can be reduced by finding more efficient heuristic strategies for performing the classical optimization of these parameters. Indeed, there are indications that this is possible: for example, instead of starting the optimization from randomly chosen initial values, as it is done in \cite{guerreschi19}, a strategy making use of observed patterns in the optimal parameters requires exponentially less time in order to achieve a similar performance \cite{zhou20}. In general, the performance of QAOA critically depends on how exactly the parameter optimization is carried out, in particular how the initial values fed into the classical optimization routine are chosen. Whereas \cite{zhou20} contains an in-depth study of the performance of QAOA on rather abstract MaxCut problems, the present paper provides a similar analysis for the use case of portfolio optimization, which constitutes one of the most discussed problems in finance \cite{mansini99, detemple14, Mansini.2015, Kalayci.2019}. 

In addition to previous works on portfolio optimization by QAOA  \cite{hodson19, Egger.2020,baker22}, the present paper therefore focuses on a number of technical issues relevant for the practical implementation of QAOA, not all of which are discussed in the present literature. By presenting the details of our implementation, which we have found to yield consistently good results, we provide a useful guide for all readers who want to apply QAOA to portfolio optimization or similar optimization problems. Furthermore, since the QAOA performance varies for different instances (i.e., different assets from which the portfolio is chosen), we present a new criterion for distinguishing between \lq easy\rq\ and \lq hard\rq\ instances of the portfolio optimization problem. Finally, viewing at the application of QAOA on error-prone hardware, we also analyze how the QAOA performance is modified by the presence of errors.

Correspondingly, the paper is organized as follows: In Sec.~\ref{sec:portopt}, we define the portfolio optimization problem as a quadratic binary optimization problem including a constraint given by the total number of assets to be included in the portfolio. In order to measure the performance of a given portfolio, we modify the standard definition of the so-called \lq approximation ratio\rq\ $r$ such that it is independent of the particular way how the constraint is implemented (see Sec.~\ref{sec:measures}). Since one such way consists of adding a penalty term, we explain our method for appropriately choosing the corresponding penalty factor $A$ (see Sec.~\ref{sec:penalty}).

The following Sec.~\ref{sec:qaoa} exposes the required concepts for solving the portfolio optimization problem using QAOA. In general, this method consists of a $p$-fold alternating application of two different operators (phase separating and mixing operators) depending on respective parameters $\vec{\gamma}=(\gamma_1,\dots,\gamma_p)$ and $\vec{\beta}=(\beta_1,\dots,\beta_p)$.
Apart from the standard implementation \cite{farhi14} (using a \lq soft constraint\rq\ based on the above-mentioned penalty term), we outline several possible modifications for incorporating the constraint in an alternative way (\lq hard constraint\rq) by using so-called XY mixing operators \cite{hadfield19} (see Sec.~\ref{sec:mixers}). Furthermore, we present in detail our strategy for performing the optimization of the QAOA parameters (see Sec.~\ref{sec:optimization}). This includes the suitable choice of a global scaling factor $\lambda$ (the relevance of which has, to our knowledge, not been recognized in the present literature) as well as four different heuristic methods for choosing appropriate initial values of the parameters $\vec{\gamma}$ and $\vec{\beta}$
for increasing depth $p$.

The results obtained by applying our algorithm to different instances of portfolio problems based on real market data (taken from the German index DAX) are presented in Sec.~\ref{sec:results}. Starting with ideal, error-free simulations (using Qiskit's statevector simulator), we show that two of the five variants of $XY$ mixers outperform the other ones by achieving the fastest convergence towards the optimal solution for increasing depth $p$ (see Sec.~\ref{sec:results_mixers}). Similar results are obtained when including statistical sampling noise due to a finite number of shots (using the qasm simulator). In the following, we compare the algorithm's performance on different problem instances (see Sec.~\ref{sec:hard_easy}). We define characteristic measures for an instance and show that the performance correlates with these. An explanation for this performance dependencies is given by analyzing the energy landscape of the respective optimization problem.

To account for the noise-affected nature of today's quantum computers, we systematically investigate the impact of errors on QAOA, thus adding a further dimension to its analysis (see Sec.~\ref{sec:noise}). Considering, both, the error-affected optimization landscapes and the final outcome of the optimization, we find, for the first time, that noise of varying strengths affects different mixers differently. We conclude that the expected noise levels on the quantum platform to be used should be taken into account when choosing a specific version of QAOA.

\section{Definition of the portfolio optimization problem}\label{sec:portopt}

Portfolio optimization is one of the classic problems of finance: How should a financial investor put together a portfolio of assets  (e.g. stocks) so that it generates the highest possible return on the one hand, but contains the lowest possible risk on the other? The solution to this problem was formulated in 1952 by H. Markowitz \cite{markowitz52} as a mathematical optimization problem. In its original form, continuous portfolio weights are allowed, i.e., it is assumed that, e.g., a stock is also tradable \lq in parts\rq. For a more realistic formulation, however, it is necessary to assume binary or at least integer portfolio weights. In the binary formulation, the portfolio weight of a stock is 1 or 0, depending on whether the stock is part of the portfolio or not. The portfolio optimization problem can then be formulated as follows: Consider the cost function

\begin{equation}
 F(z_1,z_2,\dots,z_n) = q\sum_{i,j=1}^n z_i z_j \sigma_{ij}-(1-q)\sum_{i=1}^n z_i \mu_i,\ \ z_i=0,1
\label{eq:cost-function}
\end{equation} 
which consists of the portfolio weights $z_i=0$ or $1$ of stock $i$, the covariance matrix of the stock returns $\sigma_{ij}$ and the expected return $\mu_i$. Furthermore, $n$ is the number of available assets and the factor $q$ sets the risk preference of the investor: $q=1$ would be used if the investor is fully risk-averse, i.e. he wants to choose the portfolio with the lowest risk, irrespective of the return. $q=0$ would be used in case of an aggressive investor who takes only the return of the stocks into account. In real-world applications, however, $q$ is set between 0 and 1.

Usually, an investment strategy consists in investing a fixed budget of money into a portfolio of securities. This budget constraint can be formulated in the binary optimization problem as follows:

\begin{equation}
\sum_{i=1}^n z_i=B\label{eq:constraint}
\end{equation}
The parameter $B$ defines the number of assets to be chosen for the portfolio and thereby constrains the money invested. 
In the following, we will refer to those portfolios $z_1,z_2,\dots,z_n$ which fulfill the constraint as \lq feasible\rq\ portfolios, whereas the remaining ones will be called \lq unfeasible\rq\ \cite{hodson19}.

\subsection{Covariances and returns}
\label{sec:covariances_returns}

In order to solve the portfolio optimization problem, the covariance matrix and the return vector must be provided. Here, we use the following prescription to calculate these values:  If the portfolio consists of $n$ assets, we use $m+1$ historical daily prices $p^{(i)}_{k}$ where $i=1,2,\dots,n$ and $k=0,1,\dots,m$. Out of these values we compute the daily percentage price change via 
\begin{equation}
r^{(i)}_k=\left(p^{(i)}_{k}-p^{(i)}_{k-1}\right)/p^{(i)}_{k}, 1\leq k\leq m, 
\label{eq:dailyPchange}
\end{equation}
and the annualized return of the portfolio:
\begin{equation}
\mu_i = \left[\prod_{k=1}^m \left(1+r^{(i)}_k\right)\right]^{\frac{252}{m}}
\label{eq:Return}
\end{equation}
The annualized covariance matrix is calculated with the equation
\begin{equation}
\sigma_{ij} = \frac{252}{m}\sum_{k=1}^m \left(r^{(i)}_k-\overline{r^{(i)}}\right)\left(r^{(j)}_k-\overline{r^{(j)}}\right)
\label{eq:CovarianceM}
\end{equation}
where $\overline{r^{(i)}}=\frac{1}{m}\sum_{k=1}^m r^{(i)}_k$ is the mean of the daily price changes of asset $i$.

To investigate the portfolio optimization problem for different market segments, we calculated the annualized returns and covariance matrices for the German DAX 30 (as of 1/2/2021) between 01/01/2016 and 31/12/2020 (see supplementary material).
In the following, we will investigate different instances of the portfolio optimization problem with given total number of assets $n=5$ (with budget $B=2$) or $n=10$ (with budget $B=5$). For given $n$, different instances of the problem are obtained by randomly selecting $n$ out of the 30 Dax assets. The risk preference factor will be set to $q=1/3$.

\subsection{Performance measures}
\label{sec:measures}

In order to quantify which values of the function $F$ correspond to \lq good\rq\ or \lq bad\rq\  
portfolios, we first determine the minimum and maximum value of $F$ among all feasible portfolios:
\begin{eqnarray}
F_{\rm min} & = & \min_{\sum_i z_i = B} F(z_1,\dots,z_n)\\
F_{\rm max} & = & \max_{\sum_i z_i = B} F(z_1,\dots,z_n)\label{eq:Fmax}
\end{eqnarray}
Then we define the approximation ratio of a given portfolio:
\begin{equation}
r(z_1,\dots,z_n)=\left\{\begin{array}{cc} \frac{F(z_1,\dots,z_n)-F_{\rm max}}{F_{\rm min}-F_{\rm max}} & \text{if }\sum_i z_i = B\\
0 & \text{if }\sum_i z_i \neq B\end{array}\right.
\end{equation}
Note that the optimal portfolio yields $r=1$, whereas $r=0$ corresponds to the worst feasible or an unfeasible portfolio. Let us note that this definition of the approximation ratio is independent of the particular method used to enforce the constraint (see below) \cite{baker22}. 

The QAOA algorithm returns different portfolios with different probabilities, with corresponding average approximation ratio $r$. An alternative measure for the performance of QAOA is the probability $P$ of obtaining the optimal portfolio. In contrast to the approximation ratio, however, the latter does not take into account the possibility that alternative portfolios might exhibit values of $F$ very close to $F_{\rm min}$, which would also be acceptable for the portfolio manager. 

\subsection{Penalty factor}
\label{sec:penalty}

In the standard version of QAOA (see below), the minimization is performed with respect to all states, including those which do not fulfill the budget constraint, Eq.~(\ref{eq:constraint}). In order to deal with the latter, a penalty term can be added to the function $F$ as follows:
\begin{equation}
F^{(A)}(z_1,\dots,z_n)=F(z_1,\dots,z_n)+A\left(
\sum_{i=1}^n z_i-B\right)^2\label{eq:FA}
\end{equation}
The prefactor $A$ should be chosen large enough such that, on the one hand, all unfeasible states (i.e. those which do not fulfill the constraint) yield  $F^{(A)}>F_{\rm min}$. On the other hand, if  $A$  is chosen too large, the performance of QAOA is expected to deteriorate. It has been suggested to choose $A$ such that all unfeasible states yield $F^{(A)}>F_{\rm max}$
\cite{hodson19,baker22}. As we have found, however, this choice of $A$ is unnecessarily large: after all, it is not our primary objective to separate feasible from unfeasible states, but to identify those states which exhibit values very close to $F_{\rm min}$. For this purpose, it is sufficient to choose $A$ such that 
\begin{equation}
    F_{\rm min}^{\rm (nf)}  =  \min_{\sum_i z_i\neq B} F^{(A)}(z_1,\dots,z_n)\label{eq:Fminnf}
\end{equation}
i.e., the minimum value of $F^{(A)}$ among all unfeasible states, is not \lq too close\rq\ to $F_{\rm min}$. More precisely, we use the following procedure to ensure that
\begin{equation}
F_{\rm min}^{\rm (nf)}\geq \frac{1}{2}(F_{\rm min}+\overline{F})
\label{eq:Amin}
\end{equation}
where 
\begin{equation}
\overline{F} = \frac{1}{\binom{n}{B}} \sum_{z_1+\dots+z_n=B} F(z_1,\dots,z_n)
\end{equation}
denotes the mean value of $F$ over all feasible states:
\begin{itemize}
    \item[(i)] Use a classical algorithm to determine $F_{\rm min}$
    and $\overline F$. Start with $A=0$. 
    
    \item[(ii)] Use a classical algorithm to determine $F_{\rm min}^{\rm (nf)}$, see Eq.~(\ref{eq:Fminnf}).
    
    \item[(iii)] Check whether Eq.~(\ref{eq:Amin}) is fulfilled. If yes, return the value $A$. If not, increase $A$ by $\Delta A$, where
    \begin{equation}
        \Delta A = \frac{\frac{1}{2}(F_{\rm min}+\overline{F})-F_{\rm min}^{\rm (nf)}}{\left(\sum_i z_i^*-B\right)^2}
    \end{equation}
    and $z_1^*,\dots,z_n^*$ denotes the state where the minimum in Eq.~(\ref{eq:Fminnf}) is reached. Repeat (ii) and (iii) until the condition, Eq.~(\ref{eq:Amin}), is fulfilled.
\end{itemize}
In this paper, we will only consider examples with a small number $n$ of assets (limited by the number of qubits in present NISQ devices or simulators), where $F_{\rm min}$, $F^{\rm (nf)}_{\rm min}$ and $\overline{F}$ can be exactly determined with a classical brute force search. For a larger number of assets (when a larger number of qubits will be available in the future), the brute force search will have to be replaced by an efficient classical algorithm in order to determine these quantities at least approximately. If the value of $A$ determined as described above then turns out to be too small (i.e., if the optimal solution found by the QAOA algorithm, see Sec.~\ref{sec:qaoa}, violates the constraint), the algorithm can then be re-run with a larger value of $A$.

\section{QAOA algorithm}\label{sec:qaoa}

To solve the above optimization problem on a quantum computer, we convert the function $F^{(A)}$ into a quantum operator:

\begin{equation}
\hat{F} =  \lambda~F^{(A)}\left(\frac{\hat{I}_1+\hat{Z}_1}{2},\dots,\frac{\hat{I}_n+\hat{Z}_n}{2}\right)
\label{eq:Fhat}
\end{equation}
where $\hat{I}_j$ and $\hat{Z}_j$ denote the identity and the Pauli-$\hat{Z}$ operator, respectively, acting on qubit $j$. Additionally, we scale the function by a constant factor $\lambda>0$, the significance of which will be explained below. Taking into account  Eqs.~(\ref{eq:cost-function},\ref{eq:FA}), we can write $\hat{F}$ as a sum of two-qubit and single-qubit operators (plus an irrelevant constant term $c$)
\begin{equation}
\hat{F}= \sum_{i=1}^{n-1}\sum_{j=i+1}^n W_{ij}\hat{Z}_i\hat{Z}_j-\sum_{i=1}^n w_i \hat{Z}_i + c
\end{equation}
where $W_{ij}=\frac{\lambda}{2}(q\sigma_{ij}+A)$ and $w_i=\frac{\lambda}{2} \left[(1-q)\mu_i+A (2B-n)-q \sum_{j=1}^n \sigma_{ij}\right]$.

The QAOA variational quantum circuit then generates the following quantum state $\vert\psi_{\vec{\gamma},\vec{\beta}}\rangle$ depending on the parameters $\vec{\gamma}=(\gamma_1,\dots,\gamma_p)$ and $\vec{\beta}=(\beta_1,\dots,\beta_p)$ (with number of iterations $p$, also referred to as \lq QAOA depth\rq\ in the following):
\begin{equation}
\vert\psi_{\vec{\gamma},\vec{\beta}}\rangle_M=\hat{U}_M(\beta_p)e^{-i \gamma_p \hat{F}}\dots \hat{U}_M(\beta_2)e^{-i \gamma_2 \hat{F}}\hat{U}_M(\beta_1)
e^{-i \gamma_1 \hat{F}_A}
\vert\psi_0\rangle_M\label{eq:QAOA}
\end{equation}
where the initial state $\vert\psi_0\rangle_M$ and the operator $\hat{U}_M(\beta)$ depend on the choice of the mixer $M$, see below.
Finally, all qubits are measured with respect to the standard basis in order to determine the mean value
\begin{equation}
\left< F\right>_{\vec{\gamma},\vec{\beta}} =
\langle \psi_{\vec{\gamma},\vec{\beta}}\vert
\hat{F}\vert \psi_{\vec{\gamma},\vec{\beta}}\rangle
\label{eq:qaoa_exp}
\end{equation}
This information is then passed to a classical optimization routine, which suggests new values for the parameters $\vec{\gamma}$ and $\vec{\beta}$ in order to minimize the expectation value
$\left< F\right>_{\vec{\gamma},\vec{\beta}}$.

As explained in \cite{farhi14}, this particular choice, Eq.~(\ref{eq:QAOA}), of the QAOA circuit is motivated by the principle of adiabatic quantum computing. With $\hat{U}_M(\beta)=e^{-i\beta \hat{M}}$, the circuit can be interpreted as a Trotterized time evolution transferring the eigenstate $\vert\psi_0\rangle_M$ of the mixing operator $\hat{M}$ to the desired eigenstate of the operator $\hat{F}$. For a finite depth $p$, however, the optimal choice of the angles $\vec{\beta}$ and $\vec{\gamma}$ may also amount to a non-adiabatic annealing path \cite{zhou20}.

\subsection{Mixer}\label{sec:mixers}

\subsubsection{Standard QAOA}

In the originally proposed \cite{farhi14} version of QAOA ($M={\rm standard}$), the mixing operation is realized by single-qubit rotations applied to each qubit:
\begin{equation}
\hat{U}_{\rm standard}(\beta)=e^{i\beta \sum_{i=1}^n \hat{X}_i}
\end{equation}
The corresponding initial state is
$\vert\psi_0\rangle_{\rm standard} = \vert +\dots +\rangle$,
where $\vert +\rangle = (\vert 0\rangle+\vert 1\rangle)/\sqrt{2}$. This state is the eigenstate of the mixing operator $\hat{M}_{\rm standard}=-\sum_i \hat{X}_i$ associated to its smallest eigenvalue ($-n$).

\subsubsection{XY-mixers}\label{sec:xy-mixers}

Alternatively, the initial state and the mixer can be adapted such that only states $\vert \boldsymbol{z}\rangle=\vert z_1,\dots,z_n\rangle$ fulfilling the budget constraint, see Eq.~(\ref{eq:constraint}), are populated during the algorithm \cite{hadfield19,wang20}. Correspondingly, the initial state is chosen as a uniform superposition (Dicke state~\cite{dicke_states}) 
\begin{equation}
\vert\psi_0\rangle_{M_{XY}}=\vert D^n_B\rangle=
\frac{1}{\sqrt{\binom{n}{B}}}
\sum_{\substack{i_1,\dots,i_n=0,1 \\ i_1+\dots+i_n=B}}\vert i_1\dots i_n\rangle
\end{equation}
This state is an eigenstate of the operator $(\hat{X}_i\hat{X}_j+\hat{Y}_i\hat{Y}_j)$ (for all pairs of qubits $i$ and $j$), which essentially swaps two qubits populating different states (i.e., $\vert 01\rangle\leftrightarrow \vert 10\rangle$), thereby keeping the total budget (or number of excitations) constant. Therefore, the basic ingredient of the XY-mixers is given by the following two-qubit operation:
\begin{equation}
    \hat{R}^{(XY)}_{i,j}(\beta)=e^{i\beta(\hat{X}_i\hat{X}_j+\hat{Y}_i\hat{Y}_j)}
\end{equation}
Different versions $M_{XY}$ of the $XY$-mixer are now obtained by applying this operation to different sets $S_{M_{XY}}$ of qubit pairs:
\begin{equation}
    \hat{U}_{M_{XY}}(\beta) = \prod_{(i,j)\in S_{M_{XY}}} \hat{R}^{(XY)}_{i,j}(\beta) 
    \label{eq:UMXY}
\end{equation}
We will investigate the following four versions:
\begin{itemize}
\item[(i)] {\bf Ring mixer} ($M_{XY}={\rm ring}$)

In the {\em ring mixer}, the operation $\hat{R}^{(XY)}_{i,j}(\beta)$ is applied only to neighbouring pairs of qubits, i.e., $S_{\rm ring}=\{(1,2),(2,3),\dots,(n-1,n),(n,n+1)\}$, where qubit $n+1$ is identified with qubit 1. Note that, since $[\hat{R}^{(XY)}_{i,j},\hat{R}^{(XY)}_{j,k}]\neq 0$ for mutually distinct qubits $i$, $j$ and $k$, the ordering in which the individual operations $\hat{R}^{(XY)}_{i,j}$ are applied in Eq.~(\ref{eq:UMXY}) is relevant. It is given by the order in which the corresponding qubit pairs appear in the above set $S_{\rm ring}$, i.e., first (1,2), then (2,3) etc.

\item[(ii)] {\bf Parity ring mixer} ($M_{XY}={\rm par\_ring}$)

In the {\em parity ring mixer} , this ordering is changed such as to obtain two subsets of commuting operations $\hat{R}^{(XY)}_{i,i+1}$ with odd or even $i$, respectively, i.e., $S_{\rm par\_ring}=\{(1,2),(3,4),\dots,(n_o,n_o+1),
(2,3),(4,5),\dots,(n_e,n_e+1)\}$, where $n_e$ (or $n_o$) is the largest even (or odd) number $n_e\leq n$ (or $n_o\leq n$) (and, again, qubit $n+1$ is identified with qubit 1).

\item[(iii)] {\bf Full mixer} ($M_{XY}={\rm full}$)

In the {\em full XY-mixer} ($M_{XY}={\rm full}$), the set $S_{\rm full}$ contains {\em all} pairs of qubits, where the ordering is chosen such that as many gates as possible can be performed in parallel, thus minimizing the depth of the circuit.
If $n$ is odd, the $\hat{R}^{(XY)}_{i,j}$'s are arranged into $n$ subsets of $(n-1)/2$ commuting operations, where $i+j\ {\rm mod}\ n =k$ in subset $k$. (E.g., for $n=5$: $S_{\rm full}=\{(1,5),(2,4),(2,5),(3,4),(2,1),(3,5),(3,1),(4,5),(3,2),(4,1)\}$ with subsets $\{(1,5),(2,4)\}$, $\{(2,5),(3,4)\}$ etc.) If $n$ is even, we first generate the subsets for $n-1$ as described above, and add the missing pair of qubits to each subset. (For example, for $n=6$, we add $(3,6)$ to the first subset $\{(1,5),(2,4)\}$, $(1,6)$ to the second, etc., i.e., $S_{\rm full}=\{(1,5),(2,4),(3,6),(2,5),(3,4),(1,6),...\}$.)

Together with the phase separation operator $e^{-i\gamma \hat{F}}$, a single iteration step in the full XY-mixer QAOA algorithm, see Eq.~(\ref{eq:QAOA}), reads as follows:
\begin{equation}
\hat{U}_{\rm full}(\beta)e^{-i\gamma \hat{F}}  =  \prod_{(i,j)\in S_{\rm full}}
\hat{R}^{(XY)}_{i,j}(\beta)
\prod_{(i,j)\in S_{\rm full}}e^{-i\gamma W_{ij}\hat{Z}_i\hat{Z}_j}
\prod_{i=1}^n e^{-i\gamma w_{i}\hat{Z}_i}\label{eq:XYfull}\end{equation}

\item[(iv)] {\bf Quantum alternate mixer-phase ansatz} ($M_{XY}={\rm QAMPA}$)
	
Finally, we modify the full mixer according to the recently proposed quantum alternate mixer-phase ansatz (QAMPA) \cite{larose21}. Instead of applying the operator $e^{-i\gamma \hat{F}}$ to all qubits before applying the mixer, we reorder the terms such that, for each pair of qubits, the gates corresponding to the mixer and the phase separation are contracted as follows:
\begin{equation}
\hat{U}_{\rm MP}(\beta,\gamma)  = \prod_{(i,j)\in S_{\rm full}}e^{i\beta(\hat{X}_i\hat{X}_j+\hat{Y}_i\hat{Y}_j)-i\gamma W_{ij} \hat{Z}_i\hat{Z}_j}
\prod_{i=1}^n e^{-i\gamma w_{i}\hat{Z}_i}
\label{eq:XYreshuffled}
\end{equation}
Thereby, the QAOA circuit, see Eq.~(\ref{eq:QAOA}) is modified as follows:
\begin{equation}
\vert\psi_{\vec{\gamma},\vec{\beta}}\rangle_M=\hat{U}_{MP}(\beta_p,\gamma_p)\dots \hat{U}_{MP}(\beta_2,\gamma_2) \hat{U}_{MP}(\beta_1,\gamma_1)
\vert\psi_0\rangle_{M_{XY}}\label{eq:QAMPA}
\end{equation}
As evident from Fig.~\ref{fig:circuits} in Appendix A, this reduces the number of CNOT gates by a factor 3/4. Since the CNOT gates constitute the most important source of errors in present NISQ devices, we may expect that this version is more resilient against errors than the full $XY$ mixer.

\end{itemize}

\subsection{Optimization of the QAOA parameters}
\label{sec:optimization}

The most critical step in the actual implementation of QAOA consists of choosing the classical optimization procedure used to minimize the expectation value $\left< F\right>_{\vec{\gamma},\vec{\beta}}$, see Eq.~(\ref{eq:qaoa_exp}). Before outlining our procedure, let us discuss the significance of the scaling factor $\lambda$ introduced above in Eq.~(\ref{eq:Fhat}), which, to our knowledge, has not yet been pointed out in the present literature.

\subsubsection{Global scaling factor}
\label{sec:scaling}

At first sight, the scaling $\hat{F}\to \lambda\hat{F}$ appears to have no relevant effect. Indeed, due to the phase separation operators $e^{i\gamma_j \hat{F}}$ ($j=1,\dots,p$) in Eq.~(\ref{eq:QAOA}), this factor only leads to a rescaling $\gamma_j\to \gamma_j/\lambda$ of the angles $\vec{\gamma}$, whereas the angles $\vec{\beta}$ remain unchanged. Nevertheless, we have observed that such a scaling can have an important influence on the classical optimization. Since the optimizer used for this purpose is usually imported from a standard numerical library, it, a priori, treats all the parameters $\vec{\gamma}$ and $\vec{\beta}$ on the same footing. Numerical difficulties may arise if the optimal values of the parameters $\vec{\gamma}$ differ by orders of magnitude from those of $\vec{\beta}$. In contrast, the best performance is expected if all parameters exhibit approximately the same order of magnitude. 

For this reason, our idea is to choose $\lambda$ such that the operator $\hat{F}$ exhibits a similar spectral width as the mixing operator $\hat{M}$. The latter reads $\hat{M}_{\rm standard}=-\sum_i \hat{X}_i$ (standard mixer) or $\hat{M}_{XY}=-\sum_{(i,j)\in S_{M_{XY}}} (\hat{X}_i\hat{X}_j+\hat{Y}_i\hat{Y}_j)$ ($XY$ mixers), respectively. (Note that, in the latter case  $\hat{U}_{M_{XY}}(\beta)\neq e^{-i\beta \hat{M}_{XY}}$ due to non-commuting terms. Nevertheless, we use $\hat{M}_{XY}$ for obtaining a rough estimate of a suitable $\lambda$ value). The corresponding spectral width, defined as the difference between largest and smallest eigenvalue of $\hat{M}$, is
$\Delta M=2n$ (standard, $XY$ ring and $XY$ ring\_par) or $\Delta M=n(n-1)$ ($XY$ full and $XY$ QAMPA), respectively. 

Without scaling (i.e. for $\lambda=1$), the largest eigenvalue of $\hat{F}$ is either $F_{\rm max}$ or $F^{\rm (nf)}_{\rm max}$, see Eq.~(\ref{eq:Fminnf}) with \lq $\max$\rq\  instead of \lq $\min$\rq. Since, in case of the $XY$-mixers, states which do not fulfill the constraint are not populated, we define $\Delta F=F_{\rm max}-F_{\rm min}$ as effective spectral width. In case of the standard mixer, we take into account, both, $F_{\rm max}$ and $F^{\rm (nf)}_{\rm max}$ (where the latter is typically much larger and determined by one of the few states that violate the constraint most strongly) and define $\Delta F=\left[\left(F_{\rm max}-F_{\rm min}\right)\left(F^{\rm (nf)}_{\rm max}-F_{\rm min}\right)\right]^{1/2}$. Finally, to make the spectra of $\hat{M}$ and $\hat{F}$ comparable, we set $\lambda = \Delta M/\Delta F$.

\subsubsection{Initial values of parameters for $p=1$}
\label{sec:parameter_p1}

Another important aspect concerns the initial values of $(\vec{\gamma},\vec{\beta})$ fed into the classical optimization routine. The latter usually converges to a local minimum in the vicinity of the initial point. Therefore, the quality of the solution can be enhanced by providing a good starting point which is close to the desired global minimum. In previous work \cite{zhou20}, it has been found that a suitable initial point can be extracted from the optimal values determined previously for QAOA depth $p-1$.

\begin{figure}[]
\includegraphics[width=13cm]{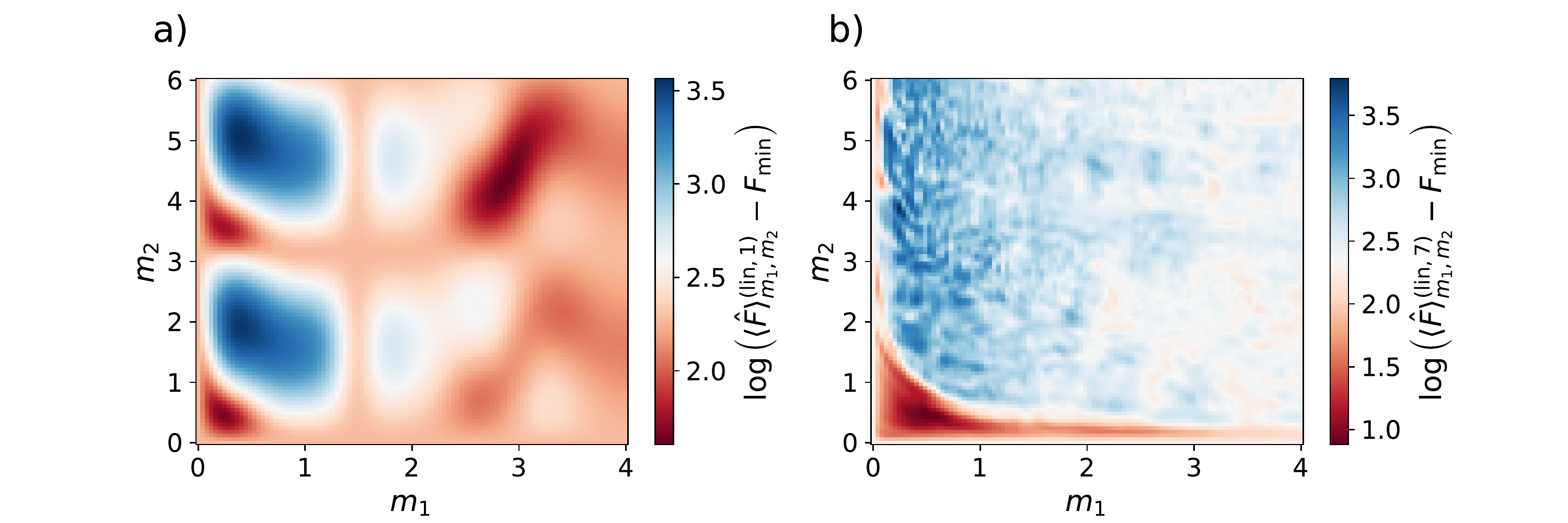}
\caption{(a) QAOA parameter landscape at $p=1$, for an exemplary portfolio optimization instance with $n=10$ assets. To enable a direct comparison with (b), the parameters $\gamma,\beta$ are expressed in terms of  $m_1=2\gamma$ and $m_2=2\beta$, respectively. The colour code shows the deviation of $\langle \hat{F}\rangle_{\gamma,\beta}=\langle \hat{F}\rangle^{({\rm lin},1)}_{m_1,m_2}$
from the optimal solution $F_{\rm min}$ on a logarithmic scale. Three pronounced minima are visible in dark red. (b) For higher depth $p=7$ with linear ansatz for the parameters $\vec{\gamma}$ and $\vec{\beta}$, see Eqs.~(\ref{eq:gammalin},\ref{eq:betalin}), only one of these minima (close to the bottom left corner) survives}
	 \label{fig:landscape}
\end{figure}

Concerning the lowest depth $p=1$, it is possible to examine a relatively large range of possible parameters $\gamma$ and $\beta$ and plot the corresponding landscape $\langle \hat{F}\rangle_{\gamma,\beta}$. Due to the periodicity of the mixers introduced above, values of $\beta$ can be restricted to the interval $[0,\pi]$, but the same is not true for $\gamma$. An example of the landscape  for an exemplary portfolio optimization instance with $n=10$ assets is shown in Fig.~\ref{fig:landscape}(a). We see several local minima, where three of them achieve similar values (two on the left, i.e., at small $m_1$, and one at $m_1\simeq 3$, $m_2\simeq 4$). As we have checked, however, not all of them are equally suitable as starting points for the extrapolation towards larger depths $p$. This can be seen by employing the following simple, linear ansatz for $(\vec{\gamma},\vec{\beta})$:
\begin{eqnarray}
    \vec{\gamma}^{({\rm lin},p)}(m_1) & = & m_1(x_1^{(p)},x_2^{(p)},\dots,x_p^{(p)})
    \label{eq:gammalin}
    \\
    \vec{\beta}^{({\rm lin},p)}(m_2) & = & m_2(1-x_1^{(p)},1-x_2^{(p)},\dots,1-x_p^{(p)})
    \label{eq:betalin}
\end{eqnarray}
where
\begin{equation}
x_i^{(p)}=\frac{2i-1}{2p}
\end{equation}
For $m_1,m_2>0$, the angles $\gamma_i^{({\rm lin},p)}(m_1)$ increase with increasing $i$ (i.e. in the same order in which they are also applied in the QAOA circuit), whereas $\beta_i^{({\rm lin},p)}(m_2)$ decrease.
This agrees with the interpretation of QAOA as quantum annealing, where $\hat{F}$ is gradually switched on while $\hat{M}$ is switched off.

If we now plot
\begin{equation}
    \langle \hat{F}\rangle^{({\rm lin},p)}_{m_1,m_2} =
    \langle \hat{F}\rangle_{{\vec{\gamma}^{({\rm lin},p)}(m_1),\vec{\beta}^{({\rm lin},p)}(m_2)}}
    \label{eq:Flin}
\end{equation}
as a function of $m_1,m_2$ for larger $p$ (e.g., $p=7$), we see that only one single pronounced local minimum remains, see Fig.~\ref{fig:landscape}(b). To find a suitable starting point for $p=1$, we therefore try 100 different combinations of parameters $m_1,m_2$ (in a logarithmically spaced grid with $\frac{1}{100} < m_1 <100$ and $\frac{\pi}{100} < m_2 < \pi$), and choose the one yielding the smallest value of $\langle \hat{F}\rangle^{\rm (lin,p_{\rm max})}_{m_1,m_2}$ (where $p_{\rm max}$ denotes the maximum depth, see below) as a starting point for the minimization of 
$\langle \hat{F}\rangle_{\gamma,\beta}$ (where $\beta=m_2/2$ and $\gamma=m_1/2$ for $p=1$), in order to arrive at the one local minimum which is suitable for extrapolation towards larger $p$.

\subsubsection{Initial values of parameters for larger $p$}
\label{sec:parameter_larger_p}

The minimization of $\langle \hat{F}\rangle$ is now performed for increasing values of $p=2,3,\dots, p_{\rm max}$. For each $p$, we apply the interpolation method described below for choosing a suitable initial point based on the optimal parameters previously found for depth $p-1$. Since we cannot strictly exclude the possibility to get stuck in a local minimum different from the global one (which is a generic problem in high-dimensional optimization), we repeat the optimization three times with additional starting points determined by different methods. In summary, we perform, for each depth $p$, four optimization runs starting from the following initial parameters:

\begin{itemize}
\item[(i)] We consider the optimized parameters 
$\vec{\gamma}^{(p-1)}=(\gamma^{(p-1)}_1,\dots,\gamma^{(p-1)}_{p-1})$ from the previous depth $p-1$ and want to determine angles
$\vec{\gamma}^{(p)}=(\gamma^{(p)}_1,\dots,\gamma^{(p)}_{p})$ as starting point for the optimization at depth $p$.
To determine $\gamma^{(p)}_i$ ($i=1,\dots,p$), we first select the two points $x^{(p-1)}_j$ and $x^{(p-1)}_{j+1}$, which are closest to $x^{(p)}_i$. Then, we define a straight line $\gamma(x)=a x +b$ such that 
$\gamma\left(x^{(p-1)}_j\right)=\gamma^{(p-1)}_j$ and $\gamma\left(x^{(p-1)}_{j+1}\right)=\gamma^{(p-1)}_{j+1}$ and choose $\gamma^{(p)}_i=\gamma\left(x^{(p)}_i\right)$. The same is repeated with the angles $\beta$.

\item[(ii)] First determine $m_1$ and $m_2$ by minimizing $\langle \hat{F}\rangle^{({\rm lin},p)}(m_1,m_2)$ (where we take $m_1$ and $m_2$ from the previous depth $p-1$ as initial point), and then use $\vec{\gamma}^{({\rm lin},p)}(m_1)$ and $\vec{\beta}^{({\rm lin},p)}(m_2)$ as initial point for minimizing $\langle \hat{F}\rangle_{\vec{\gamma},\vec{\beta}}$

\item[(iii)] The same is repeated using a quadratic ansatz
\begin{eqnarray}
    \left[\vec{\gamma}^{({\rm quad},p)}(a_1,b_1,c_1)\right]_i & = &  a_1 + b_1 x_i^{(p)} + c_1 \left(x_i^{(p)}\right)^2\\
    \left[\vec{\beta}^{({\rm quad},p)}(a_2,b_2,c_2)\right]_i & = &  a_2 + b_2 x_i^{(p)} + c_2 \left(x_i^{(p)}\right)^2
\end{eqnarray}
First, $\langle \hat{F}\rangle^{({\rm quad},p)}(a_1,b_1,c_1,a_2,b_2,c_2)= \langle \hat{F}\rangle_{{\vec{\gamma}^{({\rm quad},p)}(a_1,b_1,c_1),\vec{\beta}^{({\rm quad},p)}(a_2,b_2,c_2)}}$ is minimized to determine optimal values for the six parameters $a_1,\dots,c_2$
(where, for $p>2$, the optimized values from the previous depth $p-1$ are taken as initial values, whereas for $p=2$, we choose initial values
$(a_1,b_1,c_1,a_2,b_2,c_2)=(0,m_1,0,m2,-m2,0)$ with $m_1$ and $m_2$ as determined by the grid search described in Sec.~\ref{sec:parameter_p1}), then the corresponding angles $\vec{\gamma}^{({\rm quad},p)}$ and $\vec{\beta}^{({\rm quad},p)}$ are taken as initial point for minimizing $\langle \hat{F}\rangle_{\vec{\gamma},\vec{\beta}}$

\item[(iv)] Finally, we take the optimal angles for depth $p-1$ and add $\gamma_p=\beta_p=0$. Thereby, we recover at least the same value as for $p-1$ and ensure that $\langle \hat{F}\rangle$ monotonically decreases with increasing depth $p$.
\end{itemize}
Among these four optimization runs, we select the one achieving the smallest expectation value $\langle\hat{F}\rangle$. Before proceeding to the next depth ($p+1$), we finally rescale $\hat{F}\to \mu \hat{F}$ together with $\vec{\gamma}\to\vec{\gamma}/\mu$ (see Sec.~\ref{sec:scaling}), with $\mu$ chosen such that $\sum_i\vert\gamma_i\vert=\sum_i\vert\beta_i\vert$, to ensure that the parameters $\gamma$ and $\beta$ retain a similar order of magnitude.

\subsubsection{Classical optimizers}
\label{sec:optimizer}

Apart from trying different initial values 
$(\vec{\gamma},\vec{\beta})$ as explained above, we can choose between different routines for optimizing these values.
Concerning the simulation of the QAOA circuits, we distinguish between two different methods: on the one hand, Qiskit's \lq statevector simulator\rq\ allows us to precisely calculate the expectation value, Eq.~(\ref{eq:qaoa_exp}), without any statistical sampling error. On the other hand, the \lq qasm simulator\rq\ includes the measurement process which, for a finite number of shots (in our case: 1000), leads to a statistical sampling error due to the random measurement outcomes.

According to our experience, the gradient-based optimization routine \lq SLSQP\rq\ (which is included in the function \lq minimize\rq\ of the Python package \lq scipy.optimize\rq) yields good results in case of the statevector simulator.
The gradient-based optimization routines \lq gradient-descent\rq\ (GD) and \lq stochastic gradient
descent\rq\ (SGD) show a similar performance for the standard mixer, but perform less well than SLSQP for the XY-mixers. In case of the qasm simulator, more robust gradient-free methods (COBYLA and Nelder-Mead) are preferable, since the sampling noise prevents a precise evaluation of the gradient. 
Here, we choose the method \lq Nelder-Mead\rq\ with the following parameters: the size of the initial simplex is set to 0.5 (i.e., the simplex consists of the initial point plus $2p$ additional points obtained from the initial point by adding 0.5 to one of the $2p$ parameters) and the maximum number of iterations is limited to 10 times the number of parameters ($2p$).

\section{Results}
\label{sec:results}

\subsection{Different Mixers}
\label{sec:results_mixers}

In this section, we give a presentation and discussion of our QAOA simulation results for the 
different
mixers. 
For this purpose,
we discuss the approximation ratio and the ground state probability plotted as a function of the QAOA depth $p$ for a given number of assets that, in turn, 
coincides with
the number of qubits in the circuit. The results are generated using both
simulator methods (without and with measurement noise) with respective optimization routines as previously introduced.

For the comparison, we generate an ensemble of 20 different portfolio optimization instances with given total number $n=5$ or $n=10$ of randomly selected assets as described in Sec.~\ref{sec:covariances_returns}. We compare the convergence of approximation ratio and ground state probability of each mixer and simulation method with increasing QAOA depth $p$ (up to $p_{\rm max}=7$).

\begin{figure}[tbh]
	\centering
	\includegraphics[scale=0.5]{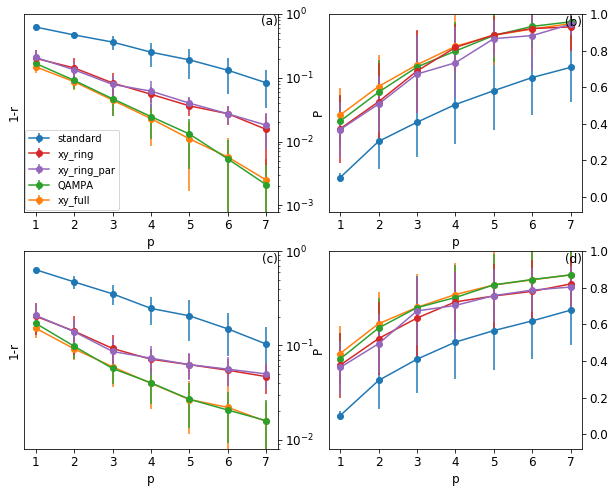}
	\caption{\small (a) Mean deviation $1-r$ of the approximation ratio from the optimal solution as a function of the QAOA depth $p$ for 20 randomly chosen portfolio optimization instances with $n=5$ assets (and budget $B=2$), using different mixers (see legend) and the statevector simulator (without measurement noise). The error bars display the standard deviation among the 20 random instances. The full XY mixer and QAMPA (which requires less CNOT gates) show the fastest convergence towards the optimal solution. (b) Mean probability $P$ of obtaining the optimal portfolio with standard deviation (error bars) as a function of QAOA depth $p$ for the statevector simulator. (c,d) Same as (a,b), but using the qasm simulator including measurement noise due to a finite number of 1000 shots instead of the statevector simulator. Significant differences with respect to the statevector simulator are only visible in the vicinity of the optimal solution ($r$ and $P$ close to 1)}
	\label{fig:appr_ratio_N5}
\end{figure}

Fig.~\ref{fig:appr_ratio_N5}~ shows the approximation ratio $r$ as well as the ground state population $P$ for 20 instances with $n=5$ assets as a function of the QAOA depth $p$ using the statevector, Fig.~\ref{fig:appr_ratio_N5}(a,b), and qasm simulator, Fig.~\ref{fig:appr_ratio_N5}(c,d). Already at $p=1$, the XY-mixers outperform the standard mixer. This can simply be explained by the fact that, in case of the XY mixers, the optimization is restricted to a smaller space \cite{Cook_2020}, since, in contrast to the standard mixer, all unfeasible portfolios (i.e., those not satisfying the budget constraint) are never populated during the algorithm (see Sec.~\ref{sec:xy-mixers}). With increasing $p$, the full and the QAMPA mixer show the fastest convergence towards the optimal solution, achieving approximation ratios $r>99\%$ without measurement noise, see Fig.~\ref{fig:appr_ratio_N5}(a). The superiority of the full XY mixer (and its variant QAMPA) as compared to the ring mixers can be traced back to their higher degree of symmetry \cite{Fuchs_2022}. Indeed, the full XY and QAMPA mixer contain all pairs of qubits, whereas only pairs of neighbouring qubits occur in the ring mixers. The full mixers therefore completely mix across all basis states at the first attempt, whereas the ring mixers take several rounds \cite{Cook_2020}. 

Concerning the probability $P$ of obtaining the optimal portfolio, see Fig.~\ref{fig:appr_ratio_N5}(b), the above discussed difference between the various mixers is less clearly visible. Taking into account the different scales (logarithmic scale for $1-r$ vs. linear scale for $P$), this is due to larger fluctuations among the 20 random instances (see error bars), such that small differences between values of $P$ close to 1 cannot be resolved. These larger fluctuations, in turn, originate from the fact that the probability $P$ depends on the population of only a single state (i.e., the optimal one), whereas the approximation ratio results from an expectation value taking into account also other feasible states, which might achieve values of the cost function close to the optimal value $F_{\rm min}$.

The results of the qasm simulator (with 1000 shots per circuit) in Fig.~\ref{fig:appr_ratio_N5}(c,d) show a  similar behavior. Only in case of the XY mixers at larger values of $p$, the statistical fluctuations due to the measurement noise lead to a slower convergence towards the optimal solution.

\begin{figure}[tbh]
	\centering
	\includegraphics[scale=0.5]{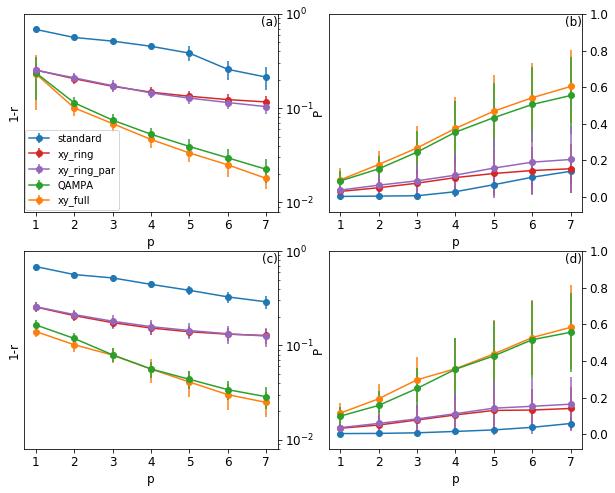}
	\caption{\small
	Same as in Fig.~\ref{fig:appr_ratio_N5} with ensemble of portfolio instances with  $n=10$ and budget $B=5$. The main conclusions, in particular concerning the superiority of the full XY and the QAMPA mixer, are similar as in Fig.~\ref{fig:appr_ratio_N5}.}
	\label{fig:appr_ratio_N10}
\end{figure}

Fig.~\ref{fig:appr_ratio_N10}~displays the results of a similar analysis with larger instances consisting of $n=10$ assets with budget $B=5$ (instead of $n=5$ and $B=2$, as in Fig.~\ref{fig:appr_ratio_N5}). Basically, they confirm our above conclusions concerning, e.g., the superiority of the full XY and the QAMPA mixer. Not surprisingly, the values of $r$ and $P$ are smaller than in Fig.~\ref{fig:appr_ratio_N5},
since the total number of possible portfolios is much larger (i.e., $10 \choose 5 = 252$ instead of $5 \choose 2=10$ feasible portfolios). Nevertheless, using one of the two best mixers (full XY or QAMPA), we still find approximation ratios as large as $r\simeq 0.98$, with corresponding probability $P\simeq 0.6$ of obtaining the optimal solution. 

\subsection{Easy and Hard Instances}
\label{sec:hard_easy}

As we have seen in the previous section, the QAOA performance, especially the probability $P$ of obtaining the optimal solution, varies a lot over a set of different instances of the portfolio optimization problem. Some instances are easier to optimize than others.
A possible cause for this performance differences could lie in the respective distributions of correlations and returns characterizing an instance.
We examine performance dependencies on these instance statistics. For this purpose, we use the variances of the returns and correlation distributions as a measure of \lq hardness\rq\ of a given instance:
\begin{align*}
 s^2_{\text{ret}} =& \text{var}\left( \{\mu_{i}  \} \right) \\
 s_{\text{cor}} ^2 =& \text{var}\left( \left \{ \frac{\sigma_{ij}}{\sqrt{\sigma_{ii}}\sqrt{\sigma_{jj}}} \mid i\leq j\right \}\right)   
\end{align*}
The correlation is normalized to a value between -1 and 1 (Pearson coefficients). This normalization resulted in more pronounced differences in statistics than using the raw correlations. 

Each instance generates an energy landscape, composed of the objective function values $F(\boldsymbol{z})$ of all possible portfolios represented by the string $\boldsymbol{z} = z_1,z_2,\dots,z_n$. We characterize the energy landscape by the mean and variance of the energy value distribution:
\begin{align*}
 \mu_{\text{energy}} =& \text{mean}\left( \{F(\boldsymbol{z})\mid \boldsymbol{z} \text{ satisfying the investment constraint} \} \right) \\
 s_{\text{energy}} ^2 =& \text{var}\left( \{F(\boldsymbol{z})\mid \boldsymbol{z} \text{ satisfying the investment constraint}  \} \right)   
\end{align*}
In the course of a Monte Carlo walk, we optimized 2400 randomly generated baskets and evaluated the performance of each instance with the measure
\begin{align*}
\text{perf}=\sqrt{G^2+R^2}.
\end{align*}
We kept the 20 best and the 20 worst performing instances and examined their statistics. 
The instances were generated as described in Sec.~\ref{sec:covariances_returns}.
We restricted the size of the instances to $n=10$ and the investment constraint to $B=5$ assets. For the optimization, we used a QAOA version with standard mixer. 
To enable a clearer comparison, we chose the same penalty factor $A=0.2$ and scaling factor $\lambda=6$ for all instances.
We further implemented \lq linear interpolation\rq\ as described in \cite{zhou20} for angle initialization and optimized up to $p=6$. The algorithm was simulated using the statevector simulator and the SLSQP method for the classical optimization step. 

\begin{figure}[h]
\subfigure{
		\includegraphics[width=0.46\textwidth]{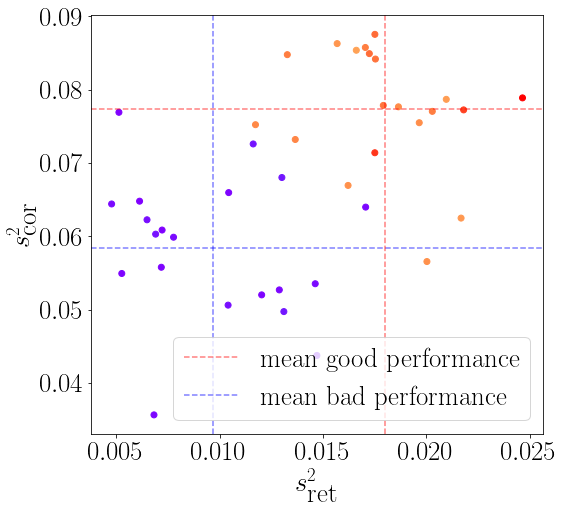}
	}
	\hfill  
\subfigure{
		\includegraphics[width=0.54\textwidth]{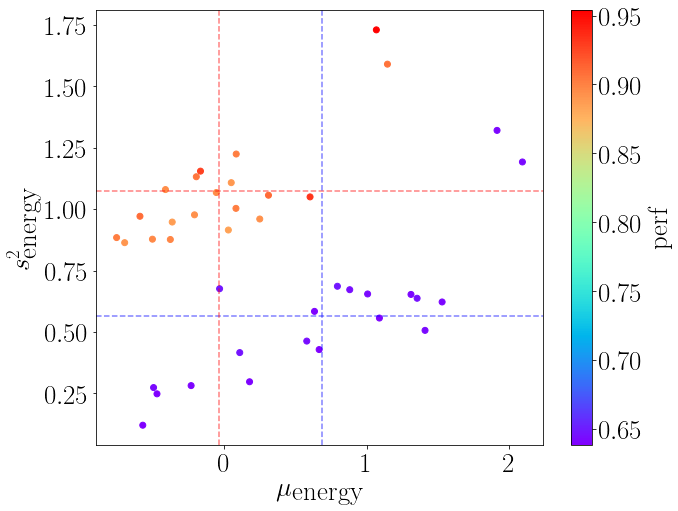}
	}

	\caption{Variance of returns $s_{\text{ret}}^2$ and normalized correlations $s_{\text{cor}}^2$ for instances of good/bad performance (left). Mean $\mu_{\text{energy}}$ and variance $s_{\text{energy}}^2$ of associated energy landscape (right)
	}
		\label{fig:instance-stats}
\end{figure}

As can be seen in Fig. \ref{fig:instance-stats}, the performance of an instance correlates with the variance of its asset correlations and returns, as well as with the mean and variance of its associated energy landscape. We repeated this procedure with historical data from MDAX 50 and found similar distributions.

These results can be interpreted as follows. Instances consisting of stocks with broadly distributed correlations are easier to optimize than those with more similar correlations. The same holds for returns. A possible reason can be found by looking at the energy landscapes created by the cost function \ref{eq:cost-function}. High variances of correlations $\sigma_{ij}$ and returns $\mu_i$ get propagated to high variance of $F(\boldsymbol{z})$ over possible $\boldsymbol{z}$. This could make the optimization step easier since the energy landscape becomes more distinct. Or, to put it figuratively: when the possible portfolios that can be built from an instance look more different in terms of correlations and returns, it is easier to choose the best one. Consequently, this dependency is an effect of the classical part of QAOA and not of the quantumness of the algorithm. It should therefore be interesting to look for this behaviour in purely classical portfolio optimization as well. Besides the characterizing statistics shown, we looked for dependencies in the eigenvalue spectra of the covariance matrices. Here, we could not find any differences for good or bad performing instances. Nevertheless, there could be other measures characterizing the instances which lead to the same, or even a more extreme performance dependency.

\subsection{Evaluation of QAOA under Noise}
\label{sec:noise}

In this section, we study the effect of noise on a given QAOA instance.
In particular, we first investigate the optimization landscape of QAOA to outline convergence issues with an increasing noise strength and then discuss the probability $P$ of obtaining the optimal solution and the approximation ratio $r$ of QAOA subject to varying noise strengths.

The evaluations in this section were performed on a QAOA instance of a portfolio optimization problem while varying the QAOA parameter $p$, the employed QAOA mixer (see section~\ref{sec:mixers}) and noise strength.
The used portfolio optimization problem consists of $n=5$ assets, i.e. five qubits, out of which $B=2$ assets must be optimally selected with a risk-return factor $q=1/3$. The corresponding covariances $\sigma_{ij}$ and returns $\mu_i$ are listed in Appendix~B.

\subsubsection{Employed Noise Model}
In order to compute QAOA subject to noise, we employ the depolarizing channel \cite{nielsen2002quantum}.
The depolarizing channel $\mathcal{E}$ acting on $n$ qubits described by a density matrix  $\rho$ is defined as
\begin{align*}
\mathcal{E}(\rho) = \frac{\eta I}{2^n}+ (1-\eta)\rho
\end{align*}
where $\eta$ is the depolarizing noise strength and $I$ denotes the identity operator.
After each single-qubit or two-qubit (CNOT) gate occurring in the QAOA ansatz circuit, the depolarizing channel is applied to the respective qubit(s). 

	\begin{figure}[tbh]
		\centering
		\includegraphics[width=1.0\textwidth]{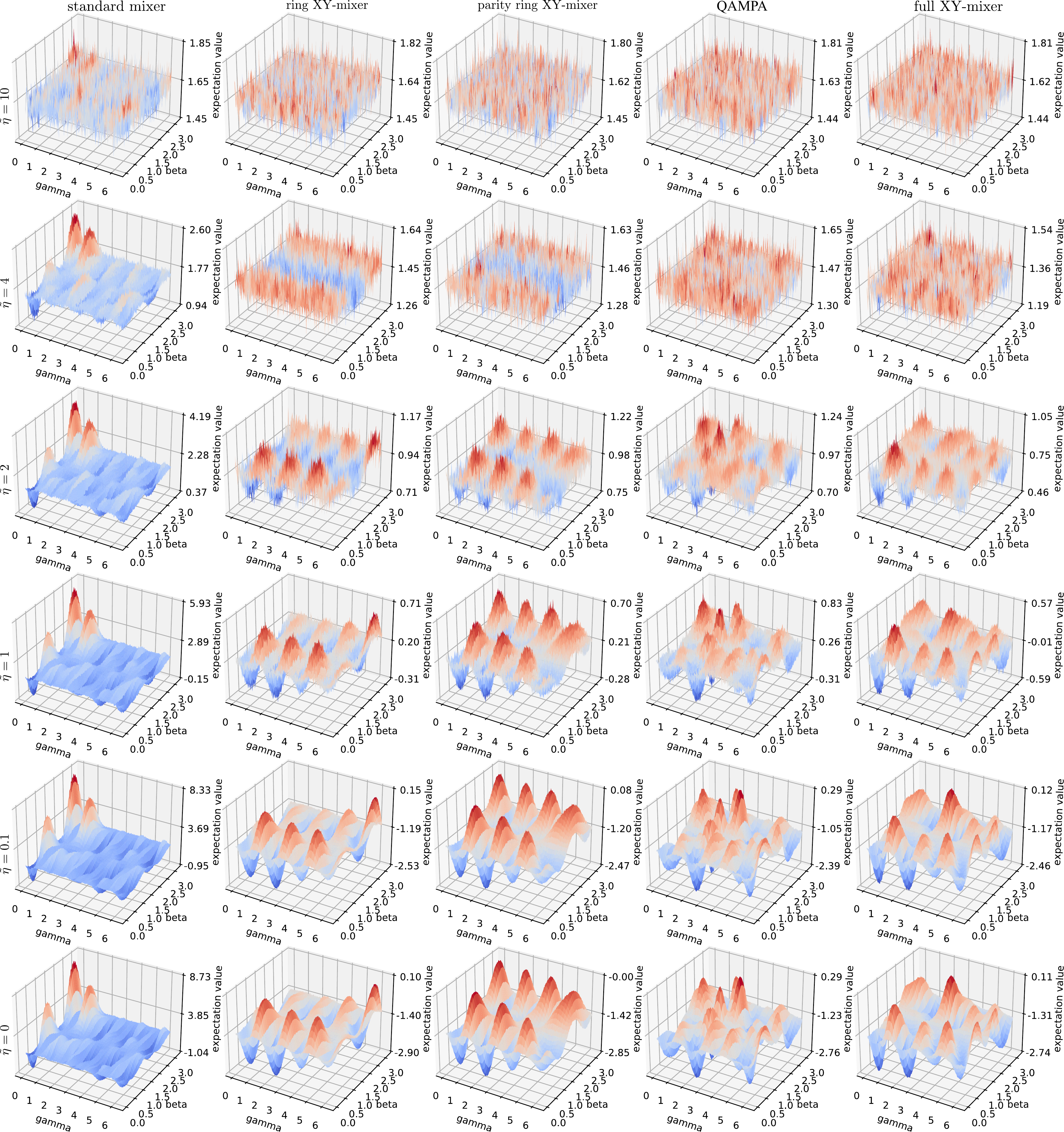}
		\caption{Expectation value of the investigated QAOA instance for the standard mixer, the ring XY-mixer, the parity ring XY-mixer, the QAMPA, and the full XY-mixer (from left to right) and noise strength $\tilde{\eta} \in\{0, 0.1, 1, 2, 4, 10 \}$ (from bottom to top), where $\tilde{\eta}$ is the noise strength normalized by the size of the quantum computation, i.e. the number of quantum gates in the QAOA ansatz circuit.
		}
		\label{fig:surface}
	\end{figure}

\subsubsection{Optimization Landscape of QAOA under Noise}

In this section, we show the optimization landscape of the investigated QAOA instance for depth $p=1$ subject to depolarizing noise for varying noise strengths and QAOA mixers.
Fig.~\ref{fig:surface}~shows the optimization landscape of the investigated QAOA instance, i.e. for a each pair $(\gamma,\beta)$ of parameters, the expectation value of the optimization function (see equation~\ref{eq:qaoa_exp}) is plotted.
The parameter $\gamma$ is varied from $0.025$ to $2\pi$ while $\beta$ is varied from $0.025$ to $\pi$ in steps of size $0.025$.
From the left to the right column, the optimization landscape of QAOA is shown for the standard mixer, the ring XY-mixer, the parity ring XY-mixer, the QAMPA and the full XY-mixer.
From the bottom to the top row, the normalized noise strength is varied as  $\tilde{\eta}\in\{0, 0.1, 1, 2, 4, 10\}$. Here, the normalized noise strength $\tilde{\eta}$ is the noise strength $\eta$ 
divided by
the size of the quantum computation, i.e. the number of quantum gates in the QAOA ansatz circuit. This normalization is used to better study the vulnerability of specific structures within different ansatz circuits to noise. Without such normalization, the performance of circuits under a constant noise level would be dominated by the circuit size, as a circuit with more gates will tend to exhibit more errors.
The $x$- and $y$-axes of the figures in Fig.~\ref{fig:surface}~show the parameters $\gamma$ and $\beta$  of QAOA, while the expectation value of the optimization function is shown on $z$-axis.

In the error-free case (bottom row of Fig.~\ref{fig:surface}), the ring XY-mixer finds the best (lowest) value of the expectation value $(-2.90)$, followed by the parity ring XY-mixer $(-2.85)$, QAMPA $(-2.76)$, full XY-mixer $(-2.74)$ and lastly the standard mixer $(-1.04)$.
However, with increasing noise strength, the order of the mixers with respect to the yielded minimal expectation value first becomes constant at $\tilde{\eta}=0.1$ and then reverts for $\tilde{\eta}=1$, such that the full XY-mixer is able to yield a lower expectation value than all other mixers.
At a noise strengths of $\tilde{\eta}=2$ and $\tilde{\eta}=4$, the standard mixer yields the best expectation value until, at $\tilde{\eta}\geq 10$, the optimization landscape is dominated by noise for all mixers.

For higher noise strengths, e.g. $\tilde{\eta}\geq 1$, the optimization landscape begins to flatten and to deteriorate for each mixer.
The expectation value tends to become less regular and more sensitive to
small changes in the parameters $\gamma$ and $\beta$.
These effects impair the ability of the classical optimization algorithms to converge to the global minimum of the expectation value, eventually leading to a noise-induced barren plateau that incurs an exponential runtime overhead~\cite{nibp}.
As demonstrated in the top rows of Fig.~\ref{fig:surface}, the optimization landscape deteriorates stronger for XY-mixers than for the standard mixer at the same noise strength $\tilde{\eta}$.

The results in Fig.~\ref{fig:surface}~demonstrate that XY-mixers yield a better expectation value compared to the standard mixer as long as the noise strength remains below some threshold.
After that threshold is exceeded, the standard mixer tends to yield a better expectation value, making the standard mixer a better option.

Overall, the optimization landscape of the standard mixer tends to be less complex, with less local optima compared to the XY-mixers. With an increasing noise strength, the deterioration of such simpler landscapes tends to be less pronounced.
This may imply a faster convergence, requiring less optimization steps, to reach the respective global minimum with the standard mixer compared to XY-mixers for higher noise levels.

\subsubsection{Evaluation of QAOA at specific noise strengths}

In this section, the investigated QAOA instance is computed subject to the depolarizing channel for different QAOA depth $p$, mixers and noise strength $\eta \in \{0, 0.001, 0.002, ..., 0.01\}$. Note that $\eta$ is not normalized, in contrast to $\tilde\eta$ employed to generate optimization landscapes in Fig.~\ref{fig:surface}. Therefore, larger circuits are more affected by noise.
We use the approximation ratio $r$ and the probability $P$ of obtaining the optimal solution (also called \lq ground state probability\rq\ in the following) as a metric to evaluate the performance of QAOA subject to noise.
A higher value of the approximation ratio or the ground state probability indicates a better performance.

	\begin{figure}[tbh]
		\centering
		\includegraphics[scale=0.5]{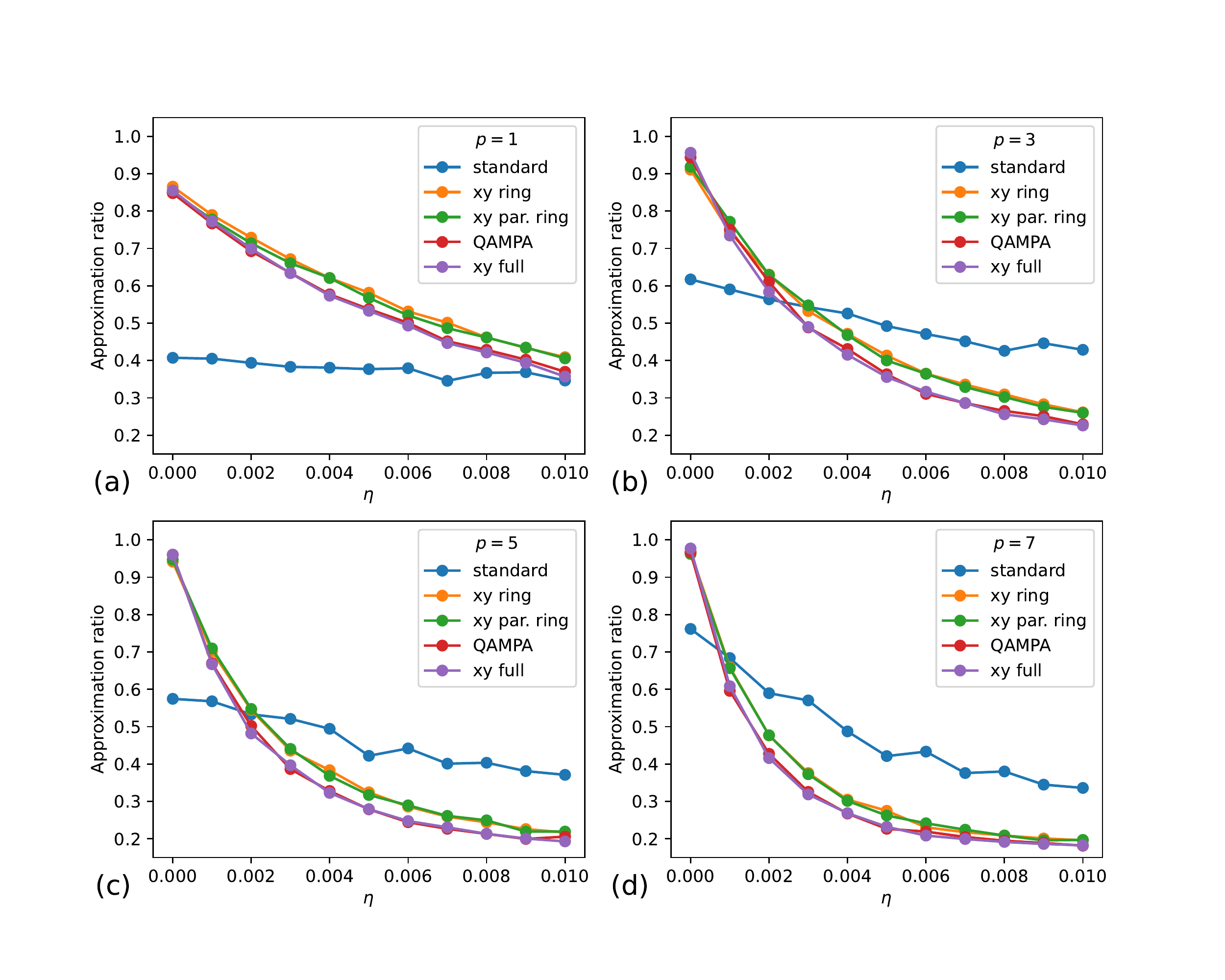}
		\caption{The approximation ratio of the investigated QAOA instance as a function of depolarizing noise strength $\eta$ for different mixers using a density matrix simulator with 8192 measurement samples. The classical optimizer COBYLA~\cite{cobyla}~was used.
		}
		\label{fig:appr_ratio_N5_dp}
	\end{figure}

Fig.~\ref{fig:appr_ratio_N5_dp}~shows the approximation ratio of QAOA for the investigated mixers and QAOA depth $p$.
For $p=1$, the XY-mixers outperform the standard mixer for all considered values of $\eta$.
For larger $p$, the standard mixer outperforms XY-mixers starting with a specific larger depolarizing noise strength, and this threshold goes down with increasing $p$ (around $0.003$, $0.002$ and $0.001$ for $p$ equal to $3$, $5$ and $7$, respectively). One reason for such behavior is the larger size of the XY-mixers, where each of the additional gates is associated with a constant noise contribution. This explanation is consistent with the dependency on $p$, where circuits for a higher $p$, both, are larger in absolute terms and have a larger size difference between standard and XY-mixers.

The ring XY-mixer and parity ring XY-mixer show better results subject to noise compared to the QAMPA and full XY-mixer at the same noise strength. This again can be explained by their smaller sizes.

\begin{figure}[tbh]
	\centering
	\includegraphics[scale=0.5]{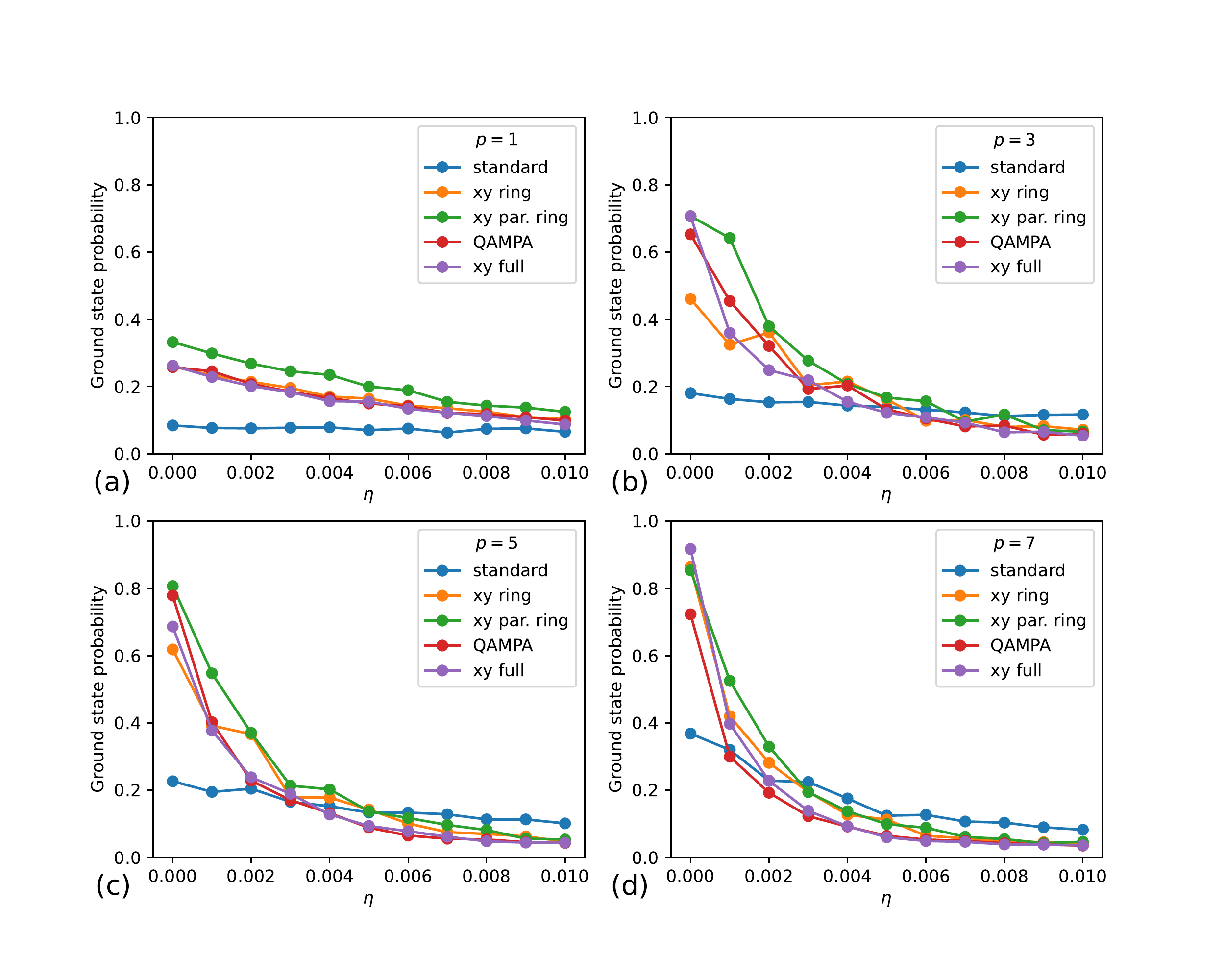}
	\caption{The ground state probability of the investigated QAOA instance as a function of depolarizing noise strength $\eta$ for different mixers using a density matrix simulator with 8192 measurement samples. The classical optimizer COBYLA~\cite{cobyla}~was used.
	}
	\label{fig:prob_N5_dp}
\end{figure}

The results for ground state probability are shown in Fig.~\ref{fig:prob_N5_dp}.
Again, for $p=1$, the parity ring XY-mixer outperforms all other mixers.
Similar to the observations for the approximation ratio, the standard mixer is able to achieve a higher ground state probability as compared to the XY-mixers for larger $p$ and higher noise strengths.
For $p=3$, $p=5$ and $p=7$, the standard mixer yields a better result than the XY-mixers at noise strengths $0.008, 0.005$, and $0.003$.
Out of the $XY$-mixers, the parity ring XY-mixer performs better subject to noise than others.

Another reason for a worsening performance of XY-mixers under higher noise levels in terms of both approximation ratio and ground state probability is their reliance on Dicke states~\cite{dicke_states} (see section~\ref{sec:mixers}). This restriction makes them more efficient in the error-free case, but noise occurring during the circuit execution can gradually destroy the Dicke state, thus violating the algorithm's assumptions. Moreover, the standard mixer requires an equal superposition as an initial state that can be prepared efficiently with a linear number of single-qubit gates, while the Dicke state needed as an input for XY-mixers must be prepared by more complex gate sequences, including multiple layers of two-qubit gates~\cite{efficient_dicke_states}.

\section{Conclusion}

The present paper contains a detailed study of portfolio optimization using different versions of the quantum approximate optimization algorithm (QAOA). 
The portfolio optimization problem is formulated as a quadratic binary optimization problem with a constraint corresponding to a given number $B$ (budget) of assets to be chosen from a list of $n$ assets (Sec.~\ref{sec:portopt}). We discuss all the technical aspects which are necessary to run different versions of the  QAOA algorithm for solving the portfolio optimization problem with good performance: first, the magnitude of the penalty factor $A$ (needed to implement the constraint in case of the standard mixer) should not be chosen larger than necessary (see Sec.~\ref{sec:penalty}). Furthermore, we propose to appropriately scale the cost function $F$ such as to align the spectra of $F$ and of the mixer $M$ with each other. Of crucial importance are the initial values of the circuit parameters $\vec{\gamma}$ and $\vec{\beta}$ fed into the classical optimization routine: similarly to \cite{zhou20}, we perform the optimization for increasing values of the QAOA depth $p$, using an extrapolation method to deduce initial values from the optimized values of $\vec{\gamma}$ and $\vec{\beta}$ obtained previously for depth $p-1$. Additionally (to avoid getting trapped in a local minimum), we perform, at each depth $p$, three additional optimization runs with initial values obtained from a linear ansatz, a quadratic ansatz, and adding zero angles (to ensure monotonicity, see Sec.~\ref{sec:parameter_larger_p}). Concerning the parameter landscape $\langle \hat{F}\rangle_{\gamma,\beta}$ at depth $p=1$, we observe several local minima and show a method for selecting one which is most suitable for extrapolation towards larger $p$ (see Sec.~\ref{sec:parameter_p1}). Finally, the proper choice of the classical optimization method depends on whether the simulation takes into account noise: in the ideal, noise-free case (using the statevector simulator), we use a gradient-based method (e.g. SLSQP), whereas more robust, gradient-free methods like Nelder-Mead or COBYLA are preferable in the presence of measurement noise due to a finite number of shots (see Sec.~\ref{sec:optimizer}). 

The performance of different mixers in solving 20 randomly chosen instances of the portfolio optimization problem with $n=5$ and $n=10$ assets is investigated in Sec.~\ref{sec:results_mixers}. Not surprisingly, the XY mixers are superior to the standard mixer, since they restrict the optimization to the subspace of \lq feasible\rq\ portfolios (i.e., to those that satisfy the budget constraint). The best results are achieved by the full XY mixer and the QAMPA (quantum alternate mixer-phase ansatz, see Sec.~\ref{sec:mixers}), which differs from the full XY mixer in the ordering of mixing and phase-separating operators applied to all pairs of qubits. For selecting $B=5$ out of $n=10$ assets with QAOA depth $p=7$, both find the optimal portfolio with probability $P\simeq 0.6$. Their better performance as compared to the two ring mixers is attributed to their higher degree of symmetry with respect to qubit permutations. Moreover, we find that the measurement noise (with 1000 shots) does not significantly reduce the performance as compared to the ideal case corresponding to an infinite number of shots, except in cases with extremely high approximation ratio (close to about 99\%).

When considering noise, the XY-mixers are in general demonstrated to be more sensitive than the standard mixer. This underscores the importance of considering the noise levels expected on the target quantum hardware when selecting the best variety of QAOA. From our specific results, purely noise-free analysis demonstrates a clear superiority of XY-mixers. However, this finding is only valid for sufficiently low noise levels, and the decision should be reverted if the actual available quantum hardware is more noisy.

Concerning the performance dependency on external factors we find clear connections with the statistics of the optimized instance, namely the variance of the correlation's and return's distributions. Broader distributions lead to a more pronounced energy landscape, simplifying the classical optimization step. Similar behaviour could be found for other optimization problems. Therefore, when benchmarking QAOA's performance the instances should be carefully picked regarding their statistics.

\backmatter

\bmhead{Supplementary information}

The files \lq DAX30\_Daily\_Covar\_5y.csv\rq\ and \lq DAX30\_Daily\_Returns\_5y.csv\rq\ attached to this submission as ancillary files contain the annualized covariance matrices and returns for the German DAX 30, calculated from data between 2016 and 2021 as described in Sec.~\ref{sec:covariances_returns}.

\bmhead{Acknowledgements}
This work is funded by the Ministry of Economic Affairs, Labour and Tourism Baden Württemberg in the frame of the Competence Center Quantum Computing Baden-Württemberg (project \lq QORA\rq).

\appendix

\section{Decomposition of mixer operators}

\begin{figure}[H]
	\includegraphics[width=12cm]{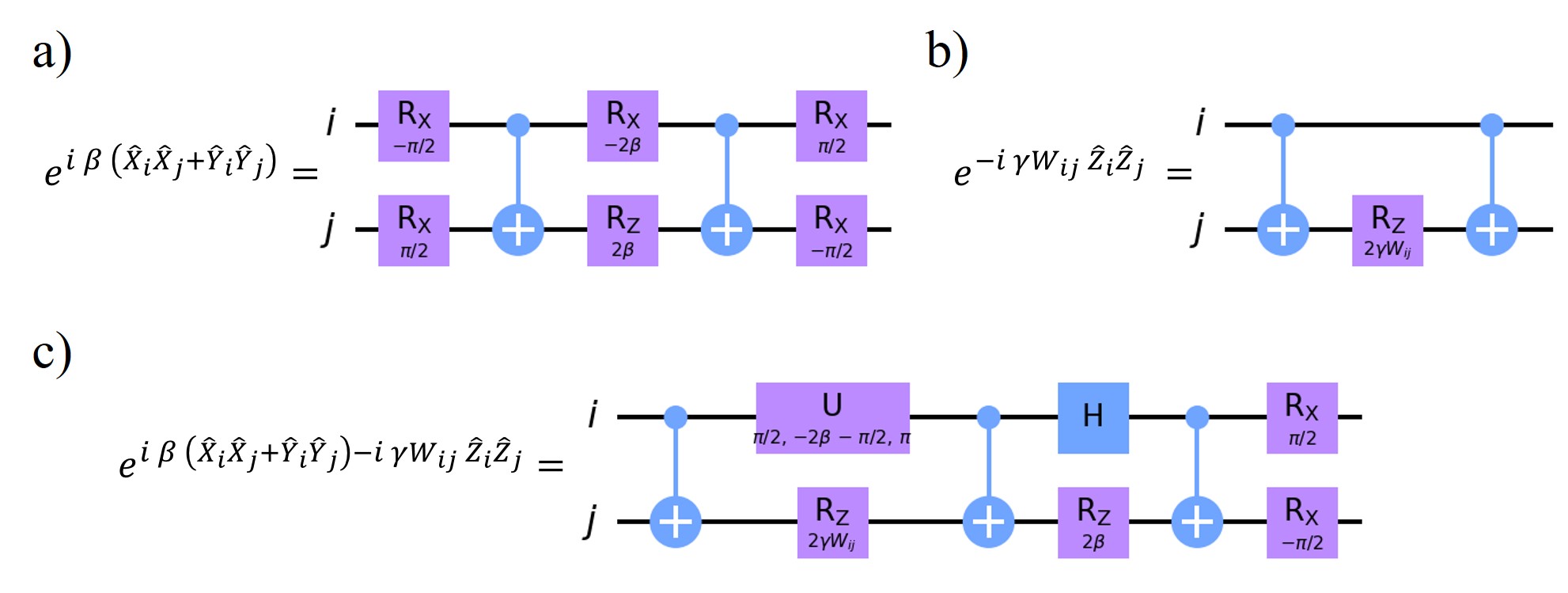}
	\caption{Quantum circuits of the two-qubit operations a) $e^{i\beta(\hat{X}_i\hat{X}_j+\hat{Y}_i\hat{Y}_j)}$ with 2 CNOT gates, b) $e^{-i\gamma W_{ij} \hat{Z}_i\hat{Z}_j}$ with 2 CNOT gates and c)  $e^{i\beta(\hat{X}_i\hat{X}_j+\hat{Y}_i\hat{Y}_j)-i\gamma W_{ij} \hat{Z}_i\hat{Z}_j}$ with 3 CNOT gates. Apart from the CNOT gates, the following single-qubit gates are used: $H$ (Hadamard gate), $R_X(\phi)=e^{-i\frac{\phi}{2} \hat{X}}$, $R_Z(\phi)=e^{-i\frac{\phi}{2} \hat{Z}}$ and the general single-qubit gate $U(\theta,\phi,\lambda)=R_Z(\phi)R_Y(\theta)R_Z(\lambda)$}
	\label{fig:circuits}
\end{figure}

Here, we show that using the reshuffled $XY$ mixer reduces the number of CNOT gates by a factor $3/4$ as compared to the full $XY$ mixer. For this purpose, Fig.~\ref{fig:circuits} shows quantum circuit representations of the two-qubit operations applied to a given pair of qubits $(i,j)$ during a single QAOA iteration step. The full $XY$ mixer, see Eq.~(\ref{eq:XYfull}), contains the circuits shown in Fig.~\ref{fig:circuits}a) and b), amounting to 4 CNOT gates for each pair of qubits. In contrast, the reshuffled $XY$ mixer, see Eq.~(\ref{eq:XYreshuffled}), only requires 3 CNOT gates, see Fig.~\ref{fig:circuits}c).

\section{Data for covariances and returns}

The covariances and returns of
the portfolio optimization instance used in Sec.~\ref{sec:noise} are provided in the tables below.

\begin{table}[h]
\begin{tabular}[h]{llccr}
LIN.DE & BAYN.DE & VNA.DE & MTX.DE & MUV2.DE\\
\hline
0.26801758 & -0.11724968 & 0.2109537 & 0.21523688 &  0.1128935 \\
\end{tabular}
\\
\caption{\label{tab:return}Return vector $\mu_{i}$ for 5 assets chosen from the German DAX 30.}
\end{table}

\begin{table}[h]
\begin{tabular}[h]{l|llccr}
 & LIN.DE & BAYN.DE & VNA.DE & MTX.DE & MUV2.DE\\
\hline
LIN.DE & 0.21117209 & 0.03030933 & 0.00941277 &  0.02972179 & 0.02922818 \\
BAYN.DE & 0.03030933 & 0.08796365 & 0.01833403 & 0.0465302 & 0.04069187 \\
VNA.DE & 0.00941277 & 0.01833403 & 0.04971719 & 0.02303918 & 0.02051608\\
MTX.DE & 0.02972179 & 0.0465302 & 0.02303918 & 0.13717214 & 0.05638483 \\
MUV2.DE & 0.02922818 & 0.04069187 & 0.02051608 & 0.05638483 & 0.06765634 \\
\end{tabular}
\\
\caption{\label{tab:covariance} Covariance matrix $\sigma_{ij}$ for 5 assets chosen from the German DAX 30.}
\end{table}





\bibliography{PortfolioOpt_QAOA}


\end{document}